# Data Acquisition System for Segmented Reactor Antineutrino Detector

Z. Hons[a,b,*], J. Vlášek[a,c,d]

[a] *Joint Institute for Nuclear Research,
Moscow Region, Dubna, Russian Federation*

[b] *NPI – Nuclear Physics Institute,
Řež, Czech Republic*

[c] *IEAP–Institute of Experimental and Applied Physics, CTU Prague
Prague, Czech Republic*

[d] *UWB–University of West Bohemia in Pilsen
Pilsen, Czech Republic*

*E-mail*: hons@ujf.cas.cz

ABSTRACT: This paper describes the data acquisition system used for data readout from a segmented detector of reactor antineutrinos with active shielding. Theoretical approach to the data acquisition is described and two possible solutions using QDCs and digitizers are discussed. Also described are the results of the DAQ performance during routine data taking operation of DANSS. DANSS (Detector of the reactor AntiNeutrino based on Solid Scintillator) is a project aiming to measure a spectrum of reactor antineutrinos using inverse beta decay (IBD) in a plastic scintillator. The detector is located close to an industrial nuclear reactor core and is covered by passive and active shielding. It is expected to have about 15000 IBD interactions per day. Light from the detector is sensed by PMT and SiPM.

---

[*] Corresponding author.

# Contents

## 1. Introduction

The DANSS project [1] aims to measure the spectrum of reactor electron antineutrinos using detection of inverse beta-decay in a plastic scintillator. Its non-flammability allows it to be located closely under a 3 $GW_{th}$ industrial nuclear power plant reactor in Kalinin, Russian Federation, where the antineutrino flux is about $10^{21}$ per second. The detector assembly has 1 $m^3$ of active volume and is covered by passive and active shielding (Figure 1). To measure the variance of antineutrino spectrum based on the distance from the reactor, the entire assembly including acquisition electronics is mounted on a lifting mechanism.

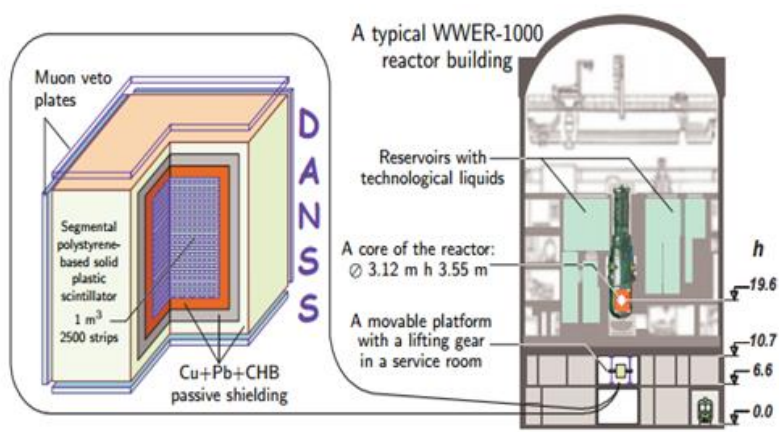

**Figure 1 Design and location of the DANSS detector**

The detector is constructed from 2500 scintillating strips with dimensions of 4×1×100 cm. Each strip is coated by a reflexive layer containing Gadolinium (Figure 2) and contains wavelength shifting fibers for light collection.

Each successive layer of the detector is built perpendicularly to the previous one. Ten parallel layers of 5 neighboring strips form one detector section (module - Figure 3). The X and Y sections are intercrossing so that the positional information of the interaction can be extracted. Every detector section is connected to one PMT.

The light is collected through wavelength shifting fibers by PMT and MPPCs. Each bundle of 100 WLS fibers from 50 strips is connected to one PMT.

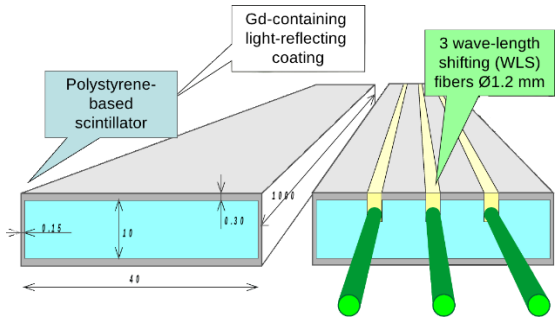

**Figure 2 Scintillating strip**

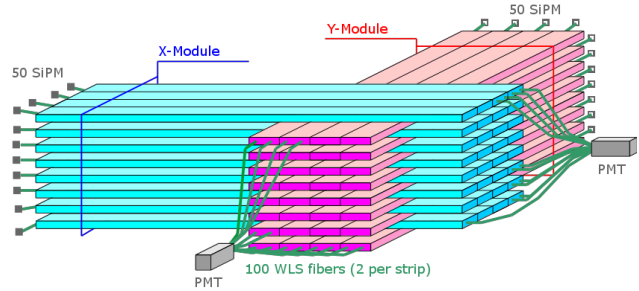

**Figure 3 Detector internal layout**

The analog output pulse from PMT analog frontend is shaped to a width of about 100 ns. An oscilloscope capture of multiple cosmic ray hits interacting with one DANSS detector section is shown in Figure 4.

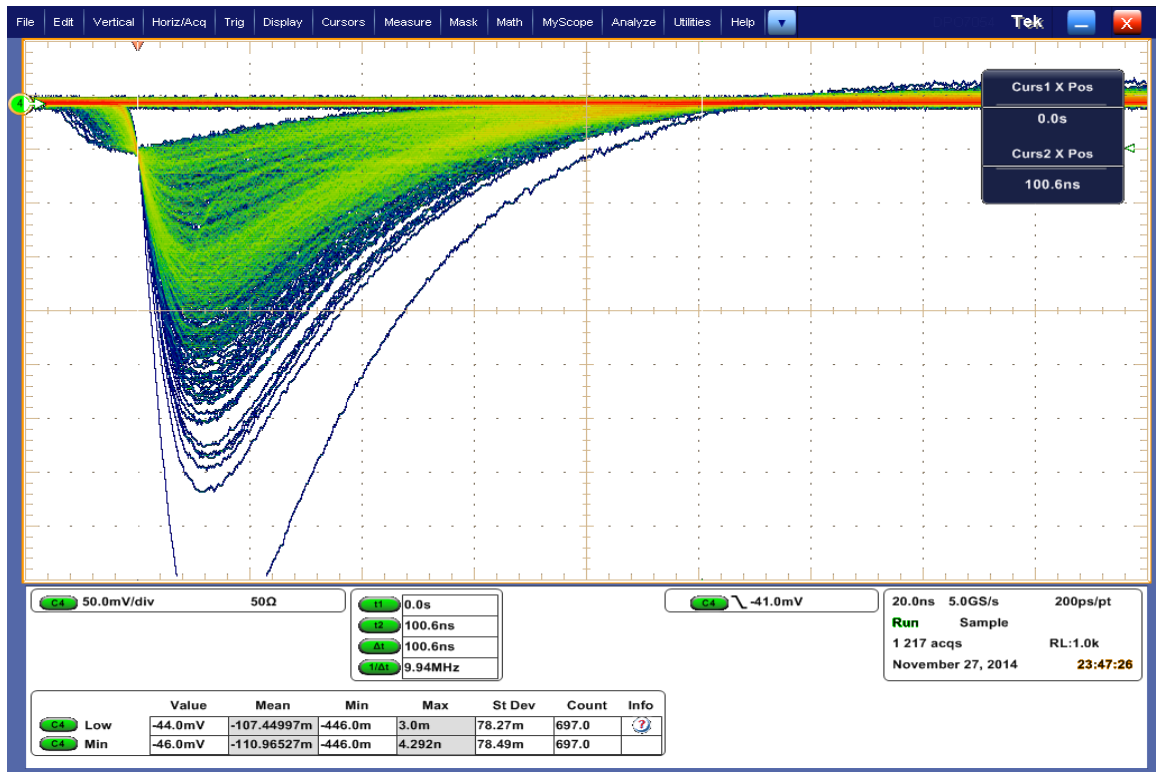

**Figure 4 Capture of cosmic background pulses from detector section**

The DANSS detector uses the inverse beta decay as its detection principle (Figure 5). An electron antineutrino coming from the reactor core interacts with a proton inside the scintillator and produces a positron and a neutron. The positron annihilates and creates a characteristic pair of 511 kEv gamma rays – "*prompt signal*". After 2 to 20 µs the neutron moderates and is captured by Gadolinium. The resulting gamma rays have a total energy of 8 MeV and should be detected within a sphere of about 20 cm from the original neutrino interaction – "*delayed signal*". This method of detection was verified by the small scale detector demonstrator DANSSino [2].

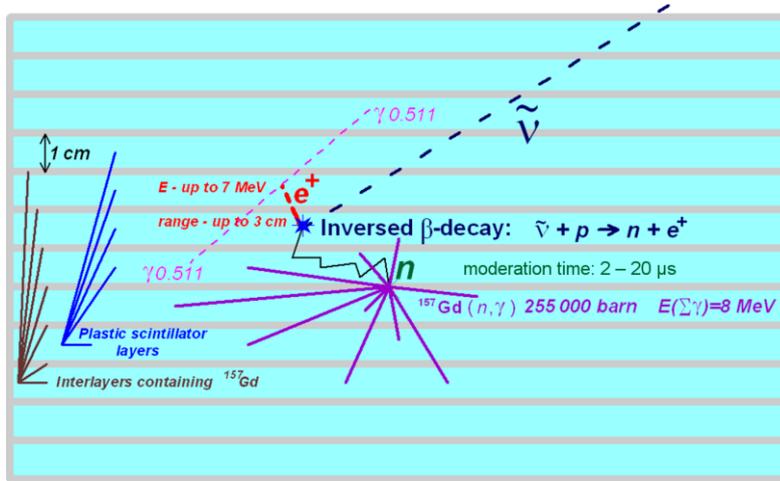

**Figure 5 Inverse beta decay detection**

Thanks to the segmented nature of the detector and the way the strips are placed, the IBD produces a characteristic signature which can be searched for. Therefore, the data acquisition system must be able to register a coincidence of two events separated by few microseconds. Generally, both prompt and delayed signals can come from the same PMT channel. The DAQ must be sufficiently responsive so that the delayed signal does not fall within the its dead time.

Additionally, the DANSS detector is using cosmic muons with vertical tracks for calibration purposes and the DAQ must also record them.

## 2. Event structure

The physics goals of the experiment and the characteristic pulse response of the detector define content and structure of relevant data event. Coincidence measurement can be done either classically when the coincidence is detected by hardware and other pulses are ignored or all detector events are registered ("event by event") and coincidences are searched for by software in the data stream based on event time stamps. Since the pulse response of the detector to particle interaction can generally create hits in multiple PMT channels (multiple *Section Pulses - SP*), the resulting composite pulse is defined as *DANSS Pulse - DP*. Since the registration of pulses is not exactly simultaneous, the first SP opens a *DANSS Window* and all SPs for a configured duration (about 50 ns) are registered as one DP. Therefore, one DP should contain all detector events related to one individual particle interaction.

In the "event by event" DAQ, the event structure contains:
- Event Timestamp
- DANSS Pulse DP
    - SP integrals (Energy)
    - Identifier of involved PMT channels

In the hardware coincidence mode, one event contains
- Event time stamp
- DP Prompt Pulse
    - SP integrals (Energy)
    - Identifier of involved PMT channels
    - Timestamp relative to Event start time
- DP Delayed Pulse
    - SP integrals (Energy)
    - Identifier of involved PMT channels
    - Timestamp relative to Event start time

The DAQ also generates events detected in the active shielding scintillators. That allows to identify cases where a cosmic muon interacted with passive shielding and caused secondary IBD-like interaction cascade.

## 3. Hardware

Conversion of electrical PMT pulse output to a number representing energy can be accomplished either using classical charge to digital converters (QDC) or using a digitizer (FADC). Both solutions have their advantages and disadvantages and since they are complementing each other the DANSS experiment has decided to use both approaches. Digitizers, working almost without a dead time, allow easy "event by event" measurement, whereas QDCs provide easily processable compact data. The problem with QDCs is that they have an inherent dead-time during which the collected charge is being converted to a number. Since the time between prompt and delayed signals of the IBD ranges from 2 to 20 μs, it is possible that the delayed signal will arrive during the dead time. In order to acquire such events, the analog signal has to be doubled and connected to a pair of QDCs and there has to be some hardware logic to route the GATE signal to a free QDC [2]. Therefore, a group of prompt QDCs and a parallel group of delayed QDCs must be used. Given the number of input channels and the necessity of adding timestamps to events a programmable logic array (FPGA) is used.

Based on the experience of the people on the team, economic conditions and previously existing solutions, it was decided to build the data acquisition system using CAEN VME modules [3]. The spectrometer is built from VME8010 crate with V2718 VME Controller optically connected to a Linux PC hosting the A2818 adapter card.

The QDC spectrometer is based on eight 16-input V965 QDCs (4 prompt, 4 delayed). These QDCs have a common gate and two internally amplified readout ranges (1× and 8×). The V965 QDCs support VME CBLT mode for significantly reduced readout time. The controlling logic is implemented in the V1495 FPGA board with 64 ECL inputs and 26 NIM outputs. The firmware generates GATE and RESET signals to appropriate QDCs. RESET is used to clear the output buffer of the QDCs when the delayed signal was not received within configured time limit. The V1495 also generates an IRQ when a coincidence condition is met – both prompt and delayed pulses were observed. Above-threshold logic pulses from the active shielding are registered by another V1495.

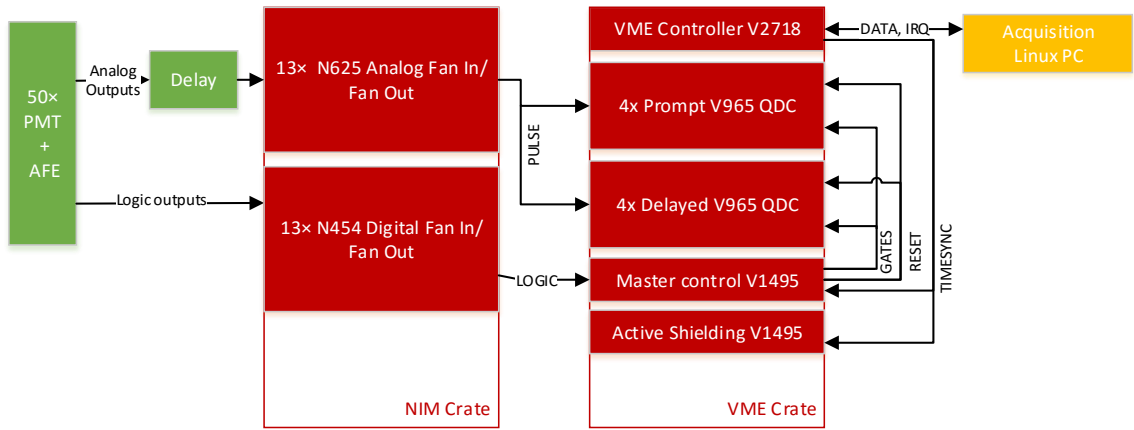

**Figure 6 Hardware schematics**

The Digitizer spectrometer uses 62.5 MSPS 64 input CAEN V1740. The sampling frequency is barely adequate to sample the pulses from DANSS analog frontend, however the obtained spectra are sufficient. The final version of the spectrometer will contain four 500 MSPS V1730 modules and will also use the optional firmware to calculate pulse integrals, reducing the recorded data volume.

Digitizers trigger internally and add timestamp to every event. 64 channels available on the V1740 will allow spectroscopic measurement of the active shielding signals.

## 4. FPGA Firmware

Charge to digital converters (QDCs) require that the gate signal comes before the analog pulse and this defines very strict timing requirements. As it is impossible to assure nanosecond timing using a PC, we chose to develop a custom firmware for a field programmable gate array (FPGA). The CAEN V1495 is a VME generic input/output card with an empty user FPGA, 64 ECL inputs, 32 ECL outputs and 3 mezzanine connectors for additional I/O modules (32×ECL, 32×LVDS, or 8× NIM). The presented system uses three NIM card configured as output.

A custom firmware samples above-threshold digital output from analog frontends using its 200 MHz system clock. Once an edge is detected on any of the 64 inputs, a short (~50 ns) window is opened which collects all the Section Pulses forming one DANSS Pulse. This DP in turn opens a much longer main coincidence window during which the system waits for a configured number of additional DPs. The main coincidence window can operate in two modes – either with fixed duration coincidence window with minimum and maximum number of pulses or a dynamic duration window which terminates as soon as enough pulses are captured. Additionally, any coincidence window with a DANSS Pulse with sufficient number of section pulses present can be deemed as successful (used for detector calibration using cosmic muon spectra). If the coincidence window does not meet configured criteria, a FAILURE pulse is generated. In case of a successful coincidence window an event is transferred to the V1495 system BLT FIFO and the DAQ is informed by asserting a VME IRQ line.

The Gate generator module routes GATE signals of configured duration to appropriate QDC and blocks further gates on the same line during a hold-off period. The window failure pulse is sent to the RESET inputs of the QDCs to clear their output buffers.

Since the V1495 does not support VME multicast addressing, the asynchronous TIMESYNC input is used to reset the system clock to synchronize multiple V1495 cards in the crate.

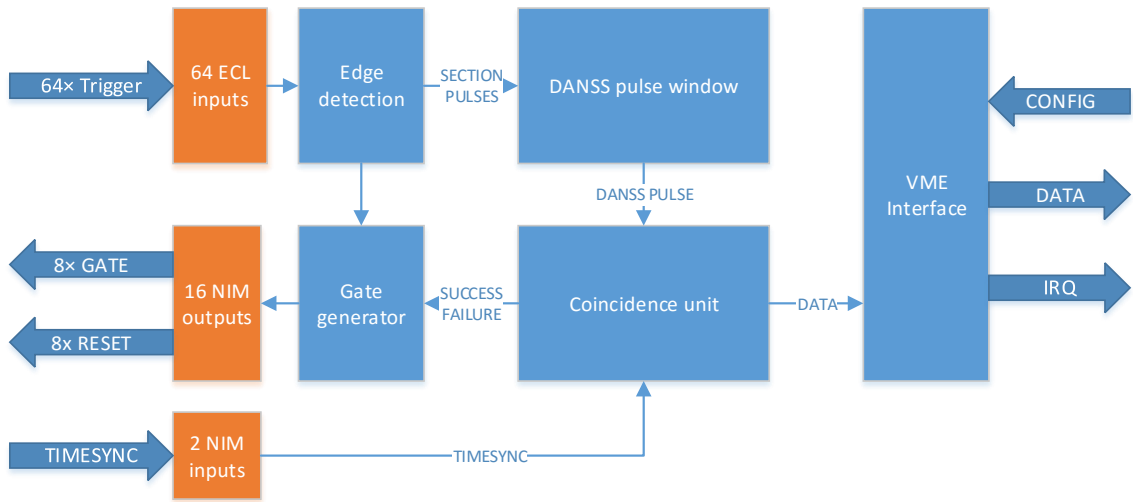

Figure 7 FPGA Firmware block diagram

## 5. Software

The spectrometer data are collected, monitored and partially on-line processed by a system described in detail in [4]. It is a system running on a Linux PC. An interesting feature of the system is that not only the vendor-specific raw data from the VME modules are written to the disk but the system also translates them into a u-data format formed by a tuple (*channel_id; value*) hiding the complexity and variability of the specific hardware. It also provides on-line calculation of areas (charges) of pulses. That allows creation of data processing programs universally usable for differently configured crates. Both raw and u-data are available for remote access via TCP/IP and so is program control. Figure 8 shows the overall DAQ schematics.

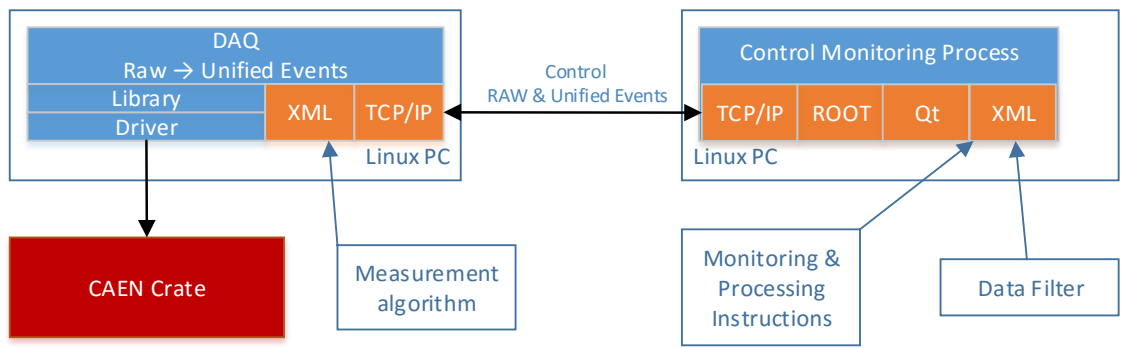

Figure 8 DAQ System Architecture

    Another important feature of the system is that the data acquisition algorithm is described using a simple programming language and is loaded from an XML configuration file. The file contains sections which relate to individual phases of data acquisition process (such as crate configuration, start of the run, IRQ handling, end of block, etc.).

    The DAQ monitoring and data processing has a very similar philosophy. The monitoring application can either receive data on-line via TCP/IP or offline from files generated by the acquisition program. The configuration file allows a simple language to define various data processing, filtering, conditional operations and finally the way how the data are presented using ROOT [5]. The system philosophy is shown in Figure 9.

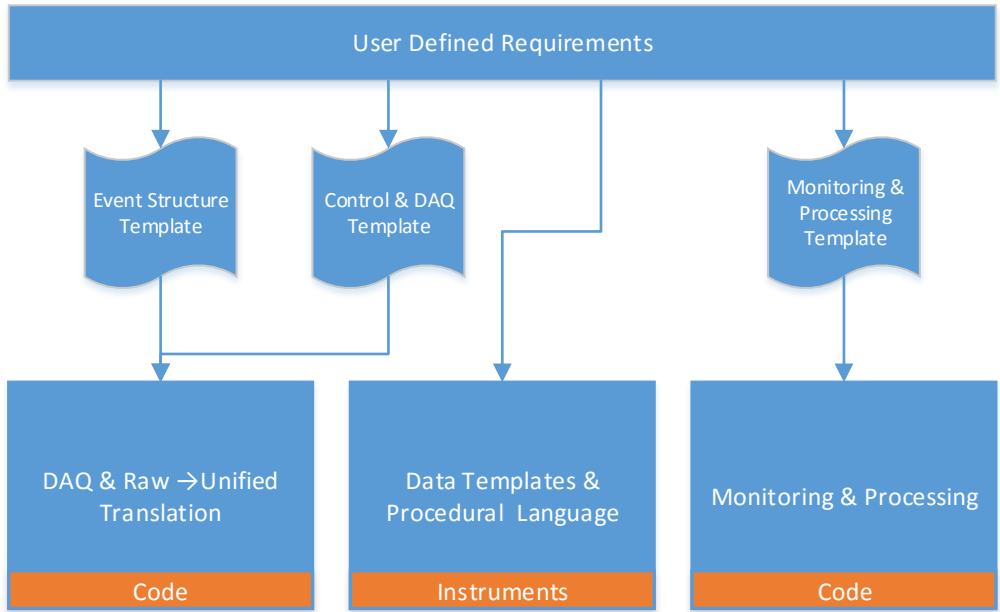

**Figure 9 DAQ and processing system framework**

## 6. System performance results

The final arrangement of the data acquisition system is shown in Figure 10. The detector, front end electronics, measurement crates and acquisition PC is located inside a technological room underneath the reactor. An Ethernet connection from this location to a monitoring room located outside the restricted access zone allows easy on-line monitoring of the performance and data retrieval.

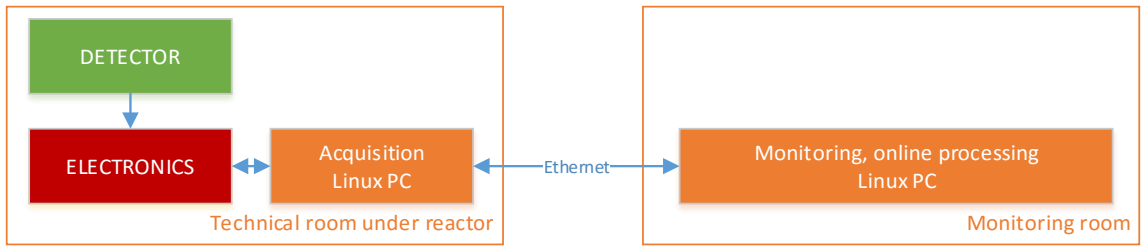

**Figure 10 Overall system schematics**

### 6.1 QDC DAQ system performance

As of spring 2016 the DANSS detector is in routine data taking operation using the QDC-based spectrometer. The measurement is done in a coincidence mode controlled by the V1495. Events from active shielding are collected by a second, independent, V1495. Due to parasitic pulses which sometimes occur approximately 500 ns after a pulse, the controlling V1495 is set to ignore any pulse closer than 2 µs. The main coincidence window lasts up to 80 µs. Data are written to disk both in the RAW and translated unified u-data formats. Every 24 hours the detector is moved into a new position. Statistical data from a single run is shown in Table 1.

| Run length | 88 020 s |
|---|---|
| Number of blocks | 1467 |
| Block length | 60 s |
| Crate events | 18 398 866 |
| Average time between IRQs | 4.8 ms |
| Average IRQ handling duration | 112 µs |
| Total handling time | 2068s |
| Total handling time [%] | 2.4% |

**Table 1 DAQ Statistics for one run**

Figure 11 shows the distribution of number of events per second. The discrete nature of the graph is due to the way this statistic is collected by the DAQ system.

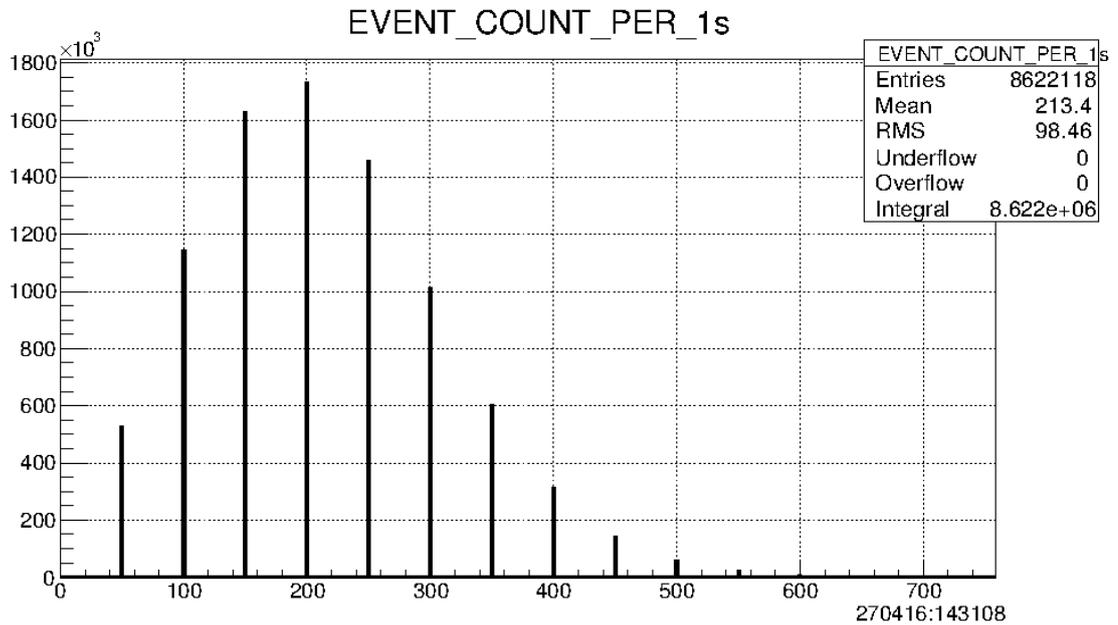

**Figure 11 Average number of events per second in one run**

Time required for crate readout is shown in Figure 12. The average time to service an IRQ is 115 µs. First, the source of the interrupt is checked and appropriate VME board status register is read. Depending on the source, either data are read from QDCs using CBLT and the control V1495, or from active shielding V1495. Because the QDC event does not contain a timestamp, the acquisition process is blocked during the readout process to assure that only related data are put into the crate event. Despite that, it can be seen that the system has been blocked only for 2.4% of the run duration.

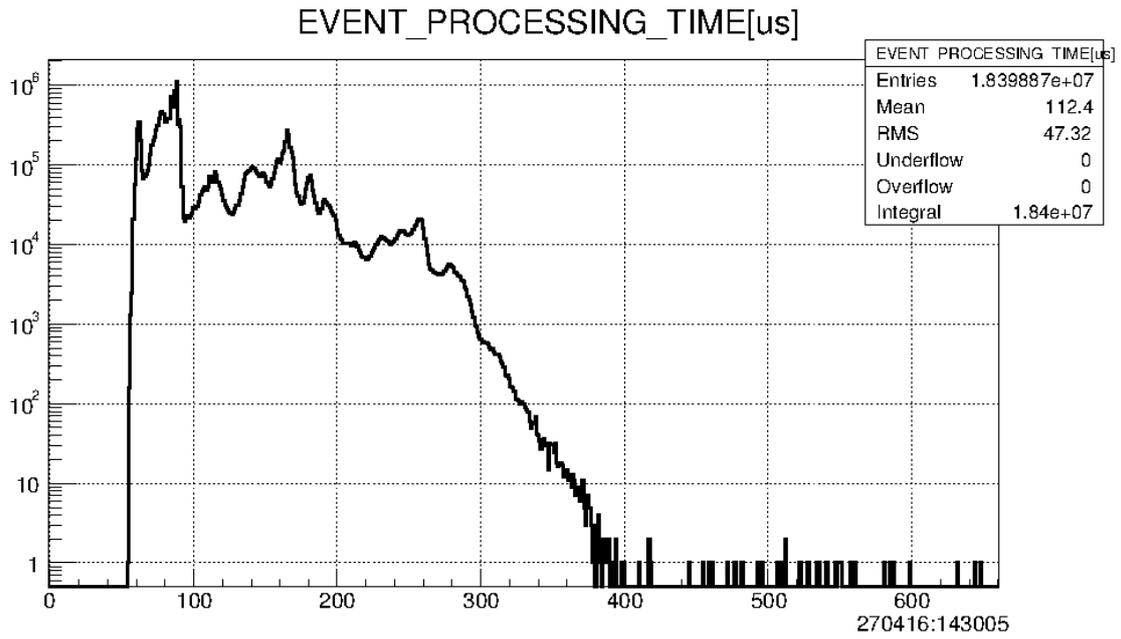

Figure 12 VME Readout time statistics

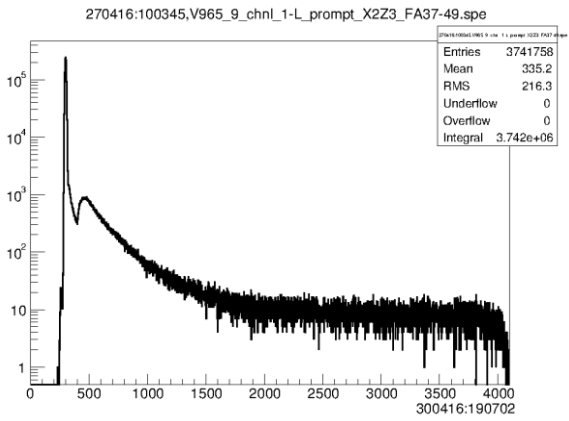  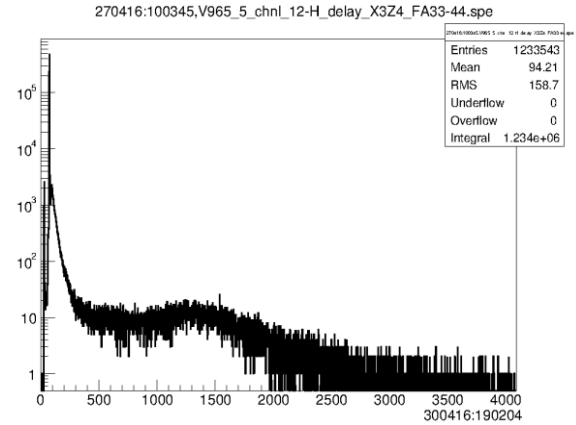

Figure 13 Prompt PMT Channel spectrum    Figure 14 Delayed PMT Channel spectrum

Figure 13 and Figure 14 show typical spectra of randomly selected PMT prompt and delayed channels during the 24 hour run.

Distribution of multiplicities of section pulses in prompt and delayed DANSS pulses and their correlation are shown in Figure 15. Both axes contain the number of section pulses in given DANSS pulses.

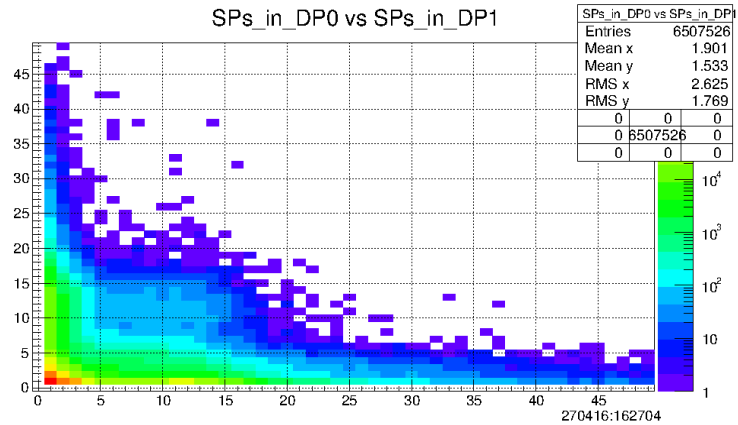

**Figure 15 Number of active PMT channels in prompt vs delayed DANSS pulses**

The spatial correlation of prompt DP and delayed DP are shown in Figure 16. X and Y axes contain the PMT channel number. These data are collected from hardware triggers by the V1495 which allows up to 64 logical inputs and only 50 PMTs are connected, explaining the gaps in the data.

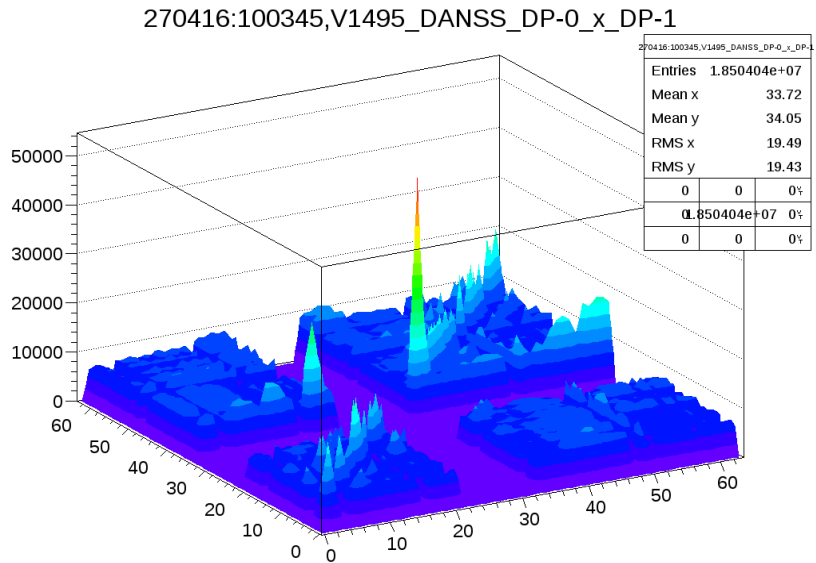

**Figure 16 Active PMT channels in prompt vs delayed pulses**

The excess of events in some channels correspond to the presence of a calibration source.

**6.2 Digitizer performance results**

Another spectrometer based on the V1740 digitizer is being connected to the detector in parallel. Many tests were performed on a small part of the detector in the laboratory. The digitizer capture length is set to 288 ns, or 18 samples. Figure 17 and Figure 18 show digitized pulses of interaction of a scintillator with cosmic muon.

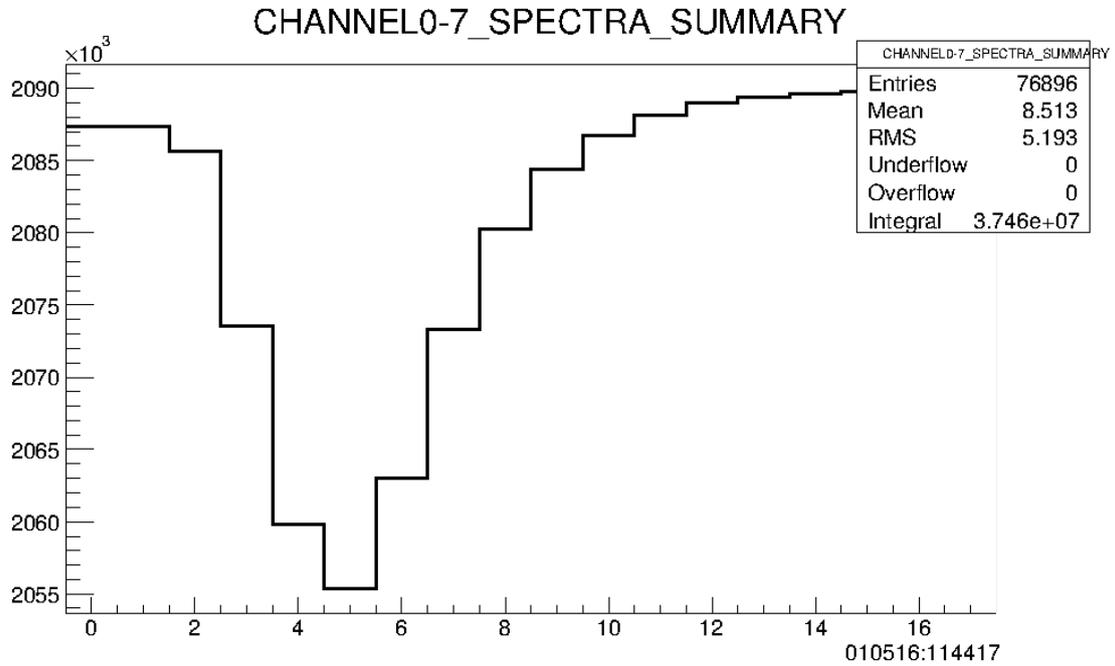

**Figure 17 Sum of all digitized waveforms**

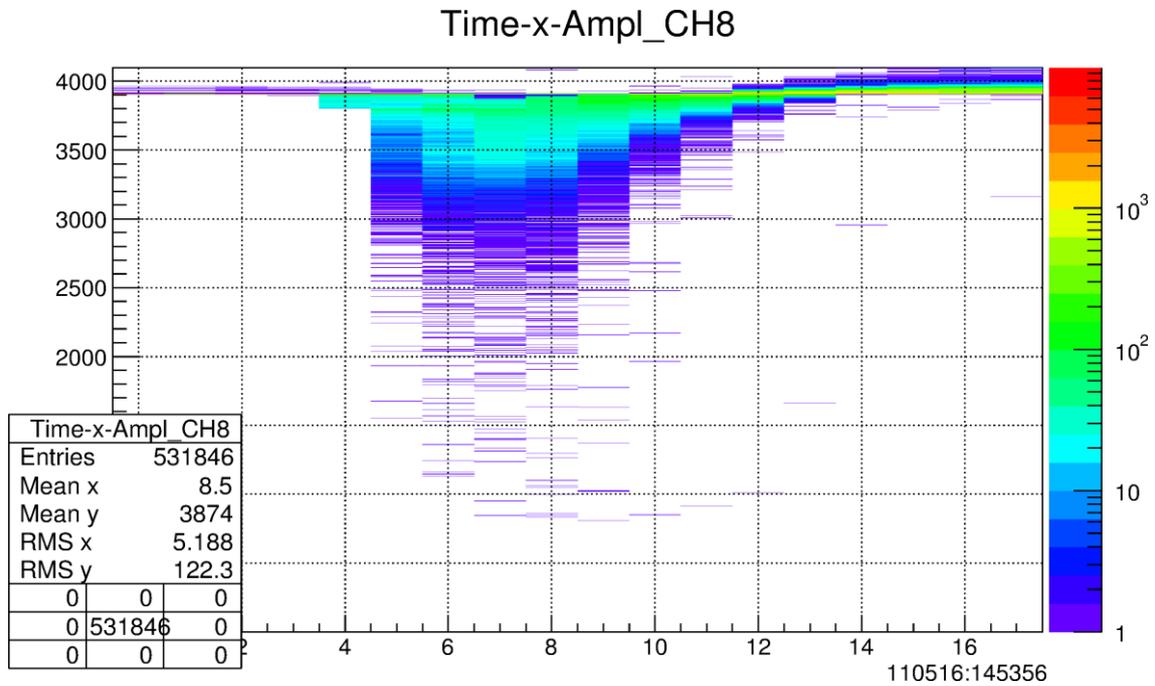

**Figure 18 Digitizer samples of natural background**

Figure 19 shows spectrum of natural background in the laboratory which was obtained using online charge integration of the digitized window and converted into the u-data format.

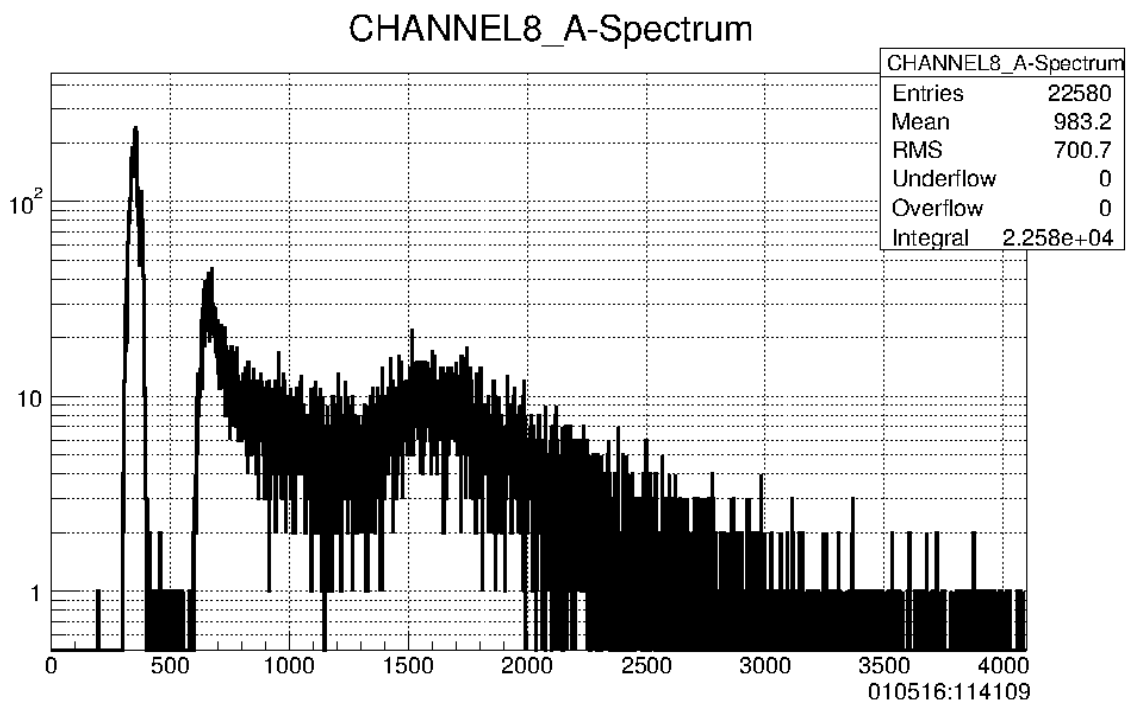

**Figure 19 Laboratory background spectrum**

The digitizer is configured with all of its 64 input channels enabled. In this case 64 * 18 samples have to be read during each IRQ. The average readout time is about 120 µs (Figure 20), thanks to the digitizer's compact event structure and support for VME block read (BLT). Like in the QDC spectrometer case, it is calculated as a difference between Linux system timestamp just after the acquisition process is woken up by IRQ and a call to wait for further interrupts. The peak around 20 µs is caused when V1740 asserts its interrupt line but the queried status register reports no available event prepared to be read out.

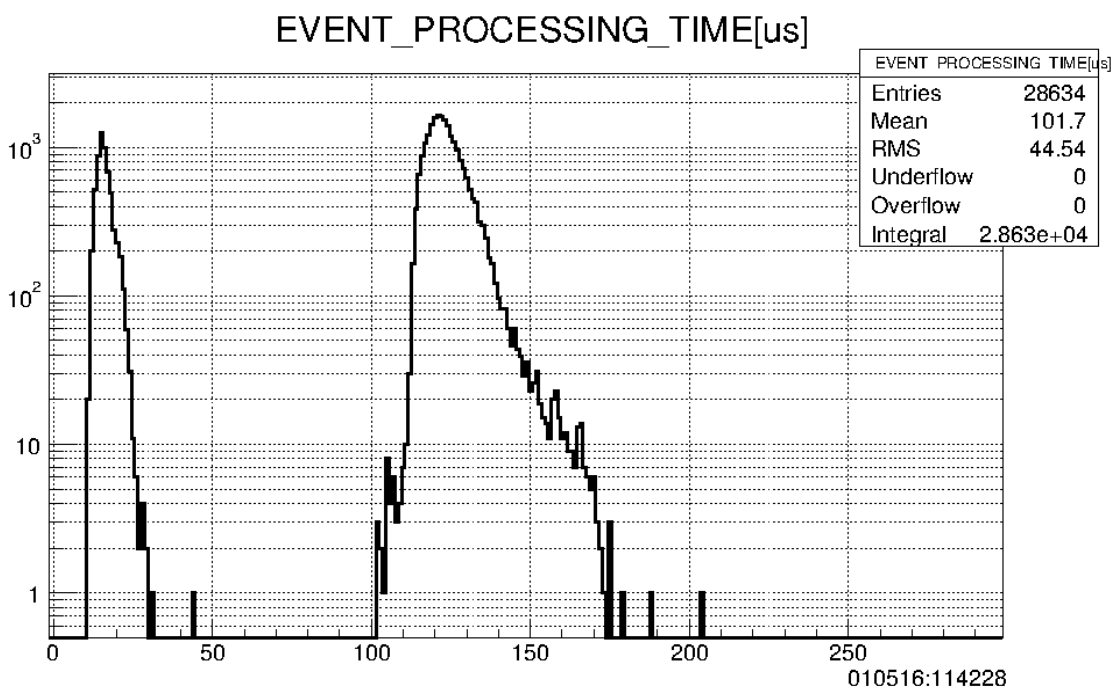

**Figure 20 Digitizer DAQ readout time**

To ensure practically zero dead time the V1740 is configured to work with 8 FIFO buffers for each channel. When the digitizer internally triggers and captures samples, it generates an interrupt and immediately is ready to capture next event. Given the average count rate of DANSS detector and its active shielding is about 2500 pulses per second and the average time it takes to read one event is about 120 μs, it is clear that it is indeed possible to capture all events in the "event by event" mode.

The V1740 is configured to not allow interleaving triggers. That means, that as soon as any channel triggers internal acquisition, other triggers are ignored and samples from all channels are recorded relative to the start of the acquisition window. Pulses which arrive later can be incomplete but the 288 ns long sampling window and its associated 50 ns DANSS window provides for correct sampling of all PMT pulses forming the DANSS Pulse and compatibility of charge integration.

## 7. Summary and conclusion

The presented system is currently routinely taking data in the Kalinin Nuclear Power Plant. It was built using commercial VME modules. Pairs of QDCs are used to reduce system dead time. We developed a custom FPGA firmware and we use our own data acquisition, monitoring and processing system. The DAQ network architecture enables us to operate the system from outside of the Power plant's radiation control zone. Fine tuning of acquisition is possible thanks to the configurable FPGA firmware and user friendly XML configuration files.

The characteristic interrupt handling time is around 120 μs, which meets the requirements of the maximal observed rate of 2500 events per second.

A parallel digitizer based DAQ is being installed in the KNPP and will start data taking during the next technological break.

**Acknowledgments**

The authors would like to thank to V. Brudanin, V. Egorov, I. Zhitnikov and other colleagues participating in the DANSS project for their many consultations and continuous support.
This work is supported by the Czech Technological Agency TE 01020445, Czech Ministry of Education Youth and Sports INGO II – LG14004.

**References**

[1]  V. Egorov and A. Starostin, "Solid scintillator detector of the reactor antineutrino DANSS," talk at TAUP2011, Munich, Germany, 2011.

[2]  V. Egorov and e. al, "DANSSino: a pilot version of the DANSS neutrino detector," p. arXiv:1305.3350, 2013.

[3]  CAEN, "Digital Pulse Processing for the Pulse Shape Discrimination," [Online]. Available: http://www.caen.it/csite/CaenProd.jsp?parent=39&idmod=770.

[4]  Z. Hons, "A versatile DAQ, monitoring and data processing system for nuclear experiments in CAMAC and VME standards," p. arXiv:1508.01379, 2015.

[5]  CERN, "ROOT Data Analysis Framework," [Online]. Available: http://root.cern.ch.

[6]  CAEN, "CAEN VME Product page," [Online]. Available: http://www.caen.it/csite/Product.jsp?parent=11&Type=Product.